\documentclass[a4paper,11pt]{article}
\pdfoutput=1 

\usepackage{jheppub} 

\usepackage[T1]{fontenc} 

\title{\boldmath Parametric Solution of a Small-Large Black Hole Coexistence Curve}

\author[a,b]{Shanshan Li,}
\author[c]{Dan-Dan Li,}
\author[a]{Li-Qin Mi}
\author[a,1]{and Zhong-Heng Li\note{Corresponding author.}}

\affiliation[a]{College of Science, Zhejiang University of
Technology, Hangzhou 310032, China}
\affiliation[b]{Fiber and Polymer Science Program, North Carolina State University, Raleigh, NC 27695, USA}
\affiliation[c]{College of Chemical Engineering, Zhejiang University of
Technology, Hangzhou 310032, China}

\emailAdd{sli31@ncsu.edu}
\emailAdd{ddli79@163.com}
\emailAdd{lqmi@zjut.edu.cn}
\emailAdd{zhli@zjut.edu.cn}

\abstract{We consider the first-order phase transition of a charged anti-de Sitter black hole, and find that the equation of state with the conditions of the two coexisting phases, leads to the two coupled equations about the thermodynamic volumes of small black hole and large black hole. By solving the equations, it is found that each reduced volume is only a function of the parameter $\omega$ . All properties of the coexistence curve can be studied from the two volume functions. In particular, each thermodynamic quantity is described by a piecewise analytic function. The demarcation point is located at $\omega_{d}=12(2\sqrt{3}-3)$. The thermodynamic function but not its derivative, is continuous at the point. This property is completely different from that of the ven der Waals fluid. Moreover, the thermodynamic behaviors as $\omega\rightarrow0$ are discussed. From which one can easily obtain some critical exponents and amplitudes for small-large black hole phase transitions.}

\keywords{Phase Transition, Parametric Solution, Black Hole}

\begin{document}
\maketitle
\flushbottom

\section{Introduction}
One of the most surprising discovery is that black holes behave as thermodynamic objects.
Original forms of the first law of black hole thermodynamics do not include a pressure term, while the mass of a black hole is interpreted as the internal energy of spacetime. However, in this case, the Smarr relation [1] is no longer satisfied for a black hole with cosmological constant.

To obtain the generalized first law, some works suggested that the cosmological constant should be considered as a thermodynamic variable [2-6]. However, to our best knowledge, no physical interpretation was ever given. In 2009, Kastor \emph{et al.} [7] proposed that the cosmological constant can be thought of as a pressure. From geometric derivations of the Smarr formula for static anti-de Sitter (AdS) black holes, the mass of an AdS black hole can naturally be interpreted as the enthalpy of the spacetime. It should be noted that an extended version of the fist law is completely consistent with the general Smarr formula [8], which obtained by scaling argument.

Of especial interest is that, in the extended phase space including pressure, the equation of state for the charged AdS black hole has the behaviour of the van der Waals (vdW) fluid [9]. This has led to numbers of investigation of black hole phase transitions from the viewpoint of traditional thermodynamics and chemistry. Two excellent review papers have been written on recent developments in this field [10,11]. However, despite extensive discussions, exact thermodynamical solutions for the black hole phase transition are still lacking. In 1982, Lekner [12] provided a parametric solution of the vdW liquid-vapor coexistence curve, which is applied to the research of various phase transtions[13-15]. Johnston [16] commented it as an elegant and very useful solution, and made a significant extension. It is found that the phase transitions of a charged AdS black hole are similar to vdW fluid, but of course they have fundamental difference. For example, in a black hole system, the Maxwell construction is effective for the pressure-thermodynamic volume plane rather the pressure-specific volume. It leads to complexity for the research of the phase transition for a charged AdS black hole. In this paper, the first-order phase transition of charged AdS black holes are discussed, and parametric solutions for a small-large black hole (SBH-LBS) coexistence curve are given. Once the analytical solution is obtained, different thermodynamic quantities can be calculated.

In the next section we discuss the behaviour of the equation of state for a charged AdS black hole, and derive the coupled equations for the thermodynamic volumes of the SBH/LBH phases from the equation of state, with the aid of the Maxwell construction and the Clausius-Clapeyron equation. In
section 3 we first introduce a new parameter $\omega$, and derive the decoupled reduced volume equation. We then give a exact solution of this equation, and show that all thermodynamic quantities in phases SBH and LBH can be obtained from the solution. In section 4 we discuss the thermodynamic behaviors near the critical point, and show that, for large $Q$, the thermodynamic quantities have similar behaviors. Finally, section 5 contains further discussion and some concluding remarks.

\section{Volume equations for SBH/LBH phase transtion}

In this section, the entropy dependence and the charge dependence of the thermodynamic volumes along a SBH-LBH coexistence curve are discussed, and then the equations of the volume including the entropy and charge are derived. We consider the proposal that, in four dimensions, thermodynamic pressure is given by $p=-\Lambda/8\pi$ , where $\Lambda$ is the cosmological constant. Then the equation of state for charged AdS black hole reads [9] (we work in the geometric units)

\begin{equation}
p=\frac{T}{v}-\frac{1}{2\pi v^{2}}+\frac{2Q^{2}}{\pi v^{4}}
,
\end{equation}
where
\begin{equation}
v=2r_{+}=2(\frac{3V}{4\pi})^{1/3}.
\end{equation}
Here $v$ and $V$ are respectively the specific volume and the thermodynamic volume, given in terms of the event horizon radius $r_{+}$, $T$ is the black hole temperature, and $Q$ its charge.

It has been shown that the equation of state (2.1) for the fixed charge $Q$ has the behaviour of the vdW fluid. According to the ordinary thermodynamic, when the two phases are in equilibrium, their pressures and chemical potentials are equal. Unfortunately, it is impossible to define the chemical potential (Gibbs function per particle) for a black hole system. Therefore, the Maxwell construction is effective for the $p-V$ plane rather the $p-v$ plane. In order to use the Maxwell construction, it is necessary to rewrite the equation of state in terms of the thermodynamic volume. Using $V=\pi v^{3}/6$, Eq.(2.1) becomes
\begin{equation}
p=\frac{1}{6}(\frac{4\pi}{3})^{1/3}[\frac{3T}{V^{1/3}}-(\frac{3}{4\pi})^{2/3}\frac{1}{V^{2/3}}+\frac{Q^{2}}{V^{4/3}}].
\end{equation}

 As similar to the $p-v$ plane, there exists a critical point:
\begin{equation}
P_{c}=\frac{1}{96\pi Q^{2}},\\
 \\ V_{c}=8\sqrt{6}\pi Q^{3}, \\
 \\ T_{c}=\frac{1}{3\sqrt{6}\pi Q},
\end{equation}
and hence $P_{c}V_{c}/T_{c}=3\pi Q^{2}/2$ (Note that is not 3/8). The isotherm that passes through the point has an inflexion there. At temperatures $T<T_{c}$, each isotherm has a maximum and a minimum, while it becomes monotonic at temperatures $T>T_{c}$. A straight horizontal segment intersecting the isotherm corresponds to a first-order SBH/LBH phase transition, which terminate at the critical point with increasing temperature, meanwhile, convert it to a second-order phase transition.

In an extended phase space, the fist law of black hole thermodynamics is given by
\begin{equation}
dM=TdS+Vdp+\Phi dQ.
\end{equation}
Here $\Phi$ and $M$ stand for the horizon electrostatic potential and black hole mass, $S$ its entropy. The Gibbs free energy can be obtained by Legendre transformation as
\begin{equation}
dG=-SdT+Vdp+\Phi dQ.
\end{equation}

As similar to the vdW fluid, the first-order SBH/LBH phase transition for fixed charge $Q$ satisfies the conditions, $T_{s}=T_{l}$, $p_{s}=p_{l}=P$, and $G_{s}=G_{l}$, where the subscripts "$s$" and "$l$" denot the thermodynamic variables in phases SBH and LBH. By integrating $dG$ along the isotherm from one intersection ($V_{s}$) with the abscissa axis to the other ($V_{l}$) as shown in Fig.1, we find
\begin{figure}
\centering
\includegraphics[width=1.2\textwidth,trim=100 320 0 320,clip]{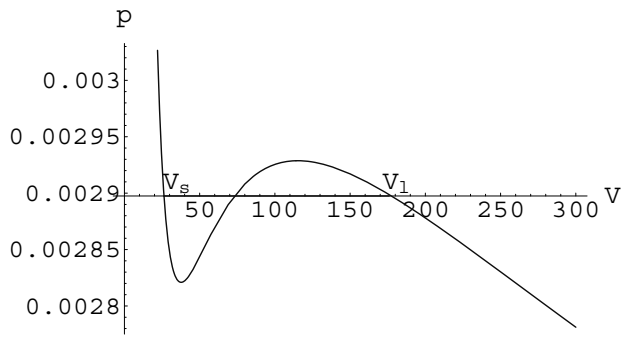}
\caption{\label{fig:i} The isotherm of a charged AdS black hole at temperature $T=0.95T_{c}$ with $Q=1$. The origin is located at $(V=0, p=P)$.}
\end{figure}

\begin{equation}
P=\frac{1}{V_{l}-V_{s}}\int_{V_{s}}^{V_{l}}pdV,
\end{equation}
which is known as the Maxwell construction. When Eq.(2.3) is substituted into Eq.(2.7) and the integration is carried out, the Maxwell construction can be written in the form
\begin{equation}
P(V_{l}-V_{s})=(\frac{\pi}{6})^{1/3}[\frac{3}{2}T(V_{l}^{2/3}-V_{s}^{2/3})-(\frac{3}{4\pi})^{2/3}(V_{l}^{1/3}-V_{s}^{1/3})-Q^{2}(V_{l}^{-1/3}-V_{s}^{-1/3})].
\end{equation}
In terms of Eq.(2.3), the condition $P=p_{s}=p_{l}$ reads
\begin{equation}
P=\frac{1}{6}(\frac{4\pi}{3})^{1/3}[\frac{3T}{V_{s}^{1/3}}-(\frac{3}{4\pi})^{2/3}\frac{1}{V_{s}^{2/3}}+\frac{Q^{2}}{V_{s}^{4/3}}]
 =\frac{1}{6}(\frac{4\pi}{3})^{1/3}[\frac{3T}{V_{l}^{1/3}}-(\frac{3}{4\pi})^{2/3}\frac{1}{V_{l}^{2/3}}+\frac{Q^{2}}{V_{l}^{4/3}}].
\end{equation}
We can use these to eliminate $P$, obtaining
\begin{equation}
T=\frac{1}{3}(\frac{1}{V_{l}^{1/3}}+\frac{1}{V_{s}^{1/3}})[(\frac{3}{4\pi})^{2/3}-Q^{2}(\frac{1}{V_{l}^{2/3}}+\frac{1}{V_{s}^{2/3}})].
\end{equation}
Equations (2.8)-(2.10) together yield the desired equation
\begin{equation}
(\frac{v_{l}}{2Q}\frac{v_{s}}{2Q}-1)^{2}-(\frac{v_{l}}{2Q}+\frac{v_{s}}{2Q})^{2}=1,
\end{equation}
where $v_{l}=2(\frac{3V_{l}}{4\pi})^{1/3}$ and $v_{s}=2(\frac{3V_{s}}{4\pi})^{1/3}$ are the specific volumes in phases SBH and LBH.

On the other hand, as shown in Fig.1, $V_{l}$ and $V_{s}$ depend on the temperature $T$. We differentiate (2.7) and have
\begin{equation}
(\frac{dp}{dT})_{Q}=\frac{1}{V_{l}-V_{s}}\int_{V_{s}}^{V_{l}}\frac{\partial p}{\partial T}dV.
\end{equation}
Compared with the standard form of the Clausius-Clapeyron equation [17]
\begin{equation}
(\frac{dp}{dT})_{Q}=\frac{\Delta S}{V_{l}-V_{s}},
\end{equation}
where $\Delta S=S_{l}-S_{s}$. From Eqs.(2.12) and (2.13), The change of entropy $\Delta S$ can be written as
\begin{equation}
\Delta S=\int_{V_{s}}^{V_{l}}\frac{\partial p}{\partial T}dV.
\end{equation}
Integration then yields
\begin{equation}
(\frac{v_{l}}{2\sqrt{\Delta S/\pi}})^{2}-(\frac{v_{s}}{2\sqrt{\Delta S/\pi}})^{2}=1.
\end{equation}

When a SBH crosses the coexistence curve and becomes a LBH, the thermodynamic volumes are not merely allowed, but  required to satisfy Eqs.(2.11) and (2.15). In this paper the equations are called the volume equations at SBH/LBH transition range.

\section{Exact solutions to volume equations}
In this section, we discuss the analytical solutions for the volume equations (2.11) and (2.15). For this purpose, we introduce a new parameter $\omega$ defined by
\begin{equation}
\omega=(\frac{\Delta S}{\pi Q^{2}})^{2}.
\end{equation}
Then the decoupled equation for the volume $\widehat{V_{s}}$ can be written in terms of the parameter $\omega$ as
\begin{equation}
1296\widehat{V_{s}}^{8/3}+432(\sqrt{\omega}-2)\widehat{V_{s}}^{6/3}+36(\omega-6\sqrt{\omega}-12)\widehat{V_{s}}^{4/3}-12(\omega+6\sqrt{\omega})\widehat{V_{s}}^{2/3}+\omega=0,
\end{equation}
where $\widehat{V_{s}}=V_{s}/V_{c}$, which is called the reduced volume. In general, the reduced variables can be obtained by dividing a quantity by its value at the critical point.

Equation (3.2) is the quartic equation with $\widehat{V_{s}}^{2/3}$. According to algebraic theory, the solution is
\begin{equation}\large
\widehat{V_{s}}^{2/3}=\left\{\begin{array}{lll}\frac{1}{12}(2-\sqrt{\omega})+\frac{1}{12\sqrt{3}}\sum ^{1}_{k=-1}[\omega+36+(-1)^{k}2\sqrt{\omega^{2}+72\omega+144}\cos\theta_{k}]^{1/2}, \\\ \textmd{if}\ \omega\leq12(2\sqrt{3}-3),\\&\\
 \frac{1}{12}(2-\omega+\frac{1}{\sqrt{3}}\sqrt{\omega+\Omega_{+}+36})+\frac{1}{6\sqrt{3}}[(\omega-\frac{\Omega_{+}}{2}+36)^{2}+\frac{3}{4}\Omega_{-}^{2}]^{1/4}\cos\frac{\theta}{2}, \\\ \textmd{if}\ \omega\geq12(2\sqrt{3}-3),
\end{array}
\right.
\end{equation}
where
\begin{equation}
\theta_{k}=k\frac{\pi}{3}+\frac{1}{3}\arccos\frac{\omega^{3}+108\omega^{2}+2160\omega-1728}{(\omega^{2}+72\omega+144)^{3/2}}, \\\ (k=-1, 0, 1),
\end{equation}
\begin{equation}
\theta=\arctan\frac{\sqrt{3}\Omega_{-}}{2(\omega+36)-\Omega_{+}},
\end{equation}
\begin{eqnarray}\large
\Omega_{\pm}&=&[\omega^{3}+108\omega^{2}+2160\omega-1728+96\sqrt{3\omega(\omega^{2}+72\omega-432)}]^{1/3}\nonumber\\&&\pm [\omega^{3}+108\omega^{2}+2160\omega-1728-96\sqrt{3\omega(\omega^{2}+72\omega-432)}]^{1/3}.
\end{eqnarray}

It should be noted that other solutions of Eq.(3.2) are unphysical. When Eqs.(2.2), (2.3) and substituted Eq.(2.15), the expression for the volume $\widehat{V_{l}}=(V_{l}/V_{c})^{2/3}$ can be obtained,
\begin{equation}
\widehat{V_{l}}^{2/3}=\widehat{V_{s}}^{2/3}+\frac{1}{6}\sqrt{\omega}.
\end{equation}

\begin{figure}
\centering 
\includegraphics[width=1.2\textwidth,trim=100 320 0 300,clip]{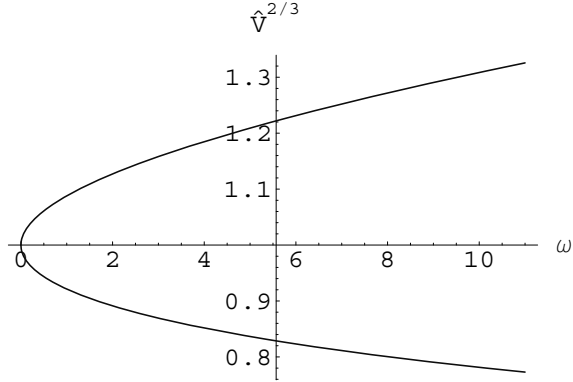}
\caption{\label{fig:i}The behaviour of $\widehat{V_{l}}^{2/3}$ and $\widehat{V_{s}}^{2/3}$ as a function of parameter $\omega$. The curve above the abscissa axis is for $\widehat{V_{l}}^{2/3}$, while the one below is for $\widehat{V_{s}}^{2/3}$. The origin is located at $(\omega=24\sqrt{3}-36, \widehat{V}^{2/3}=1)$.}
\end{figure}

From Eqs.(3.3) and (3.7), it is clear that the reduced volumes in phases SBH and LBH are determined only by the parameter $\omega=(\Delta S/\pi Q^{2})^{2}$. Interestingly enough, there exists a demarcation point, $\omega_{d}=12(2\sqrt{3}-3)$, and each reduced volume  has different functions on both sides of the demarcation point. As in Fig.2, the curve in the different quadrants corresponds to the different functions. In particular, a reduced volume is continuous at $\omega_{d}$, but not its derivative.

Note that expressions (3.3) and (3.7) are exact, and all other thermodynamic quantities in phases SBH and LBH can be derived from these two fundamentals.

\section{Thermodynamic behaviors near the critical point}

The physical connotation of the critical point is very rich. The phase equilibrium curve terminates at the critical point in the $p-T$ plane, while the SBH and LBH coexisting states become identical there, etc. So it is very important to investigate the thermodynamic behaviors near the critical point. In this section, we give the approximation expressions of some thermodynamic functions near the critical point.

When the black hole system passes the critical point, the parameter $\omega$ vanishes (due to $\Delta S=0$). Using Eqs.(2.4), (2.9), (2.10), (3.3), (3.7), and the Taylor expansion about $\omega=0$, we find the approximation values of the reduced variables, $\widehat{P}$, $\widehat{T}$, $\widehat{V_{s}}$ and $\widehat{V_{l}}$ near the critical point as follows:
\begin{equation}
\widehat{P}=\frac{P}{P_{c}}=1-\frac{1}{432}\omega+\frac{5}{373248}\omega^{2}+O(\omega^{3}),
\end{equation}
\begin{equation}
\widehat{T}=\frac{T}{T_{c}}=1-\frac{1}{1152}\omega+\frac{11}{2654208}\omega^{2}+O(\omega^{3}),
\end{equation}
\begin{equation}
\widehat{V_{s}}=\frac{V_{s}}{V_{c}}=1-\frac{1}{8}\omega^{1/2}+\frac{11}{1152}\omega+O(\omega^{3/2}),
\end{equation}
\begin{equation}
\widehat{V_{l}}=\frac{V_{l}}{V_{c}}=1+\frac{1}{8}\omega^{1/2}+\frac{11}{1152}\omega+O(\omega^{3/2}).
\end{equation}

Similarly, one can write the Clausius-Clapeyron equation (2.13) in terms of the reduced variables as
\begin{equation}
(\frac{d\widehat{P}}{d\widehat{T}})_{Q}=\frac{8}{3}-\frac{7}{1296}\omega+\frac{299}{8957952}\omega^{2}+O(\omega^{3}).
\end{equation}

On the other hand, Ref.[18] introduced the number density, which is defined as $n=1/v$, to investigate the microscopic structure of charged AdS black hole phase transitions, When one black hole phase changes into  another, the number density also suffers a sudden change. Of interest are the ratio of the LBH and SBH reduced densities
\begin{equation}
\frac{\widehat{n_{l}}}{\widehat{n_{s}}}=(\frac{V_{s}}{V_{l}})^{1/3}=1-\frac{1}{12}\omega^{1/2}+\frac{1}{288}\omega+O(\omega^{3/2}),
\end{equation}
the reduced density average,
\begin{equation}
<\widehat{n}>=\frac{1}{2}(\widehat{n_{l}}+\widehat{n_{s}})=\frac{1}{2}(\widehat{V_{l}}^{-1/3}+\widehat{V_{s}}^{-1/3})=1+\frac{1}{3456}\omega-\frac{37}{23887872}\omega^{2}+O(\omega^{3}),
\end{equation}
and the  reduced density difference,
\begin{equation}
\Delta \widehat{n}=\widehat{n_{s}}-\widehat{n_{l}}=\frac{1}{12}\omega^{1/2}-\frac{1}{4608}\omega^{3/2}+\frac{125}{95551488}\omega^{5/2}+O(\omega^{7/2}).
\end{equation}

Equations (4.1)-(4.8) described the thermodynamic behaviors as $\omega\rightarrow0$. For which, there are two cases: $\Delta S\rightarrow0$ and $Q\rightarrow\infty$. This shows that the behaviors of some reduced variables for large $Q$ are the same as for near the critical point.

\section{Discussion and conclusion}

We have investigated the parametric  solution of a SBH/LBH coexistence curve. When a SBH crosses the coexistence in the $p-T$ plane and becomes a LBH, the volumes satisfy Eqs.(2.11) and (2.15), which have elegant algebraic structures. Actually Eq.(2.15) confirms what we expected: the entropy change is one quarter of the area difference of the LBH and SBH.

The physical solutions of Eqs.(2.11) and (2.15) are given by expressions (3.3) and (3.7). All properties of the coexistence curve in terms of the parameter $\omega=(\Delta S/\pi Q^{2})^{2}$ can be studied from the expressions. It should be noted that each thermodynamic quantity is described by a piecewise analytic function with the demarcation point locates at $\omega_{d}=12(2\sqrt{3}-3)$. The physical interpretation of it is still an open question.

We have also considered the thermodynamic behaviors as $\omega\rightarrow0$. From Eqs.(4.1)-(4.8), one can easily obtain some crtical exponents and amplitudes for SBH-LBH phase transitions. For example, $\Delta \widehat{n}\rightarrow2\sqrt{2}(1-\widehat{T})^{1/2}$, which shows that the $\beta$ exponent and $b$ amplitude, defined by [16] $\Delta \widehat{n}=b(1-\widehat{T})^{\beta}$, take the values $1/2$ and $2\sqrt{2}$, respectively. The exponent is exactly the same as the vdW fluid, but the amplitude is different. These properties for near the critical point are same as for large $Q$.
\\\\{\bf
Acknowledgments} \\\\This work was supported by the National Natural
Science Foundation of China under Grants No. 11475148 and No.
11075141.



\begin{thebibliography}{99}


\bibitem{1}
L. Smarr , \emph{Mass formula for Kerr black holes}, \emph{Phys. Rev. Lett.} {\bf 30} (1973) 71.

\bibitem{2}
M. Henneaux and C. Teitelboim, \emph{The cosmological constant as a canonical variable}, \emph{Phys. Lett.} {\bf B 143} (1984) 415.

\bibitem{3}
C. Teitelboim, \emph{The cosmological constant as a thermodynamic black hole parameter}, \emph{Phys. Lett.} {\bf B 158} (1985) 293.

\bibitem{4}
M. Henneaux, and C. Teitelboim, \emph{Asymptotically anti-de Sitter spaces}, \emph{Commun. Math.} Phys. {\bf 98} (1985) 391.

\bibitem{5}
M. Henneaux and C. Teitelboim, \emph{The cosmological constant and general covariance}, \emph{Phys. Lett.} {\bf B 222} (1989) 195.

\bibitem{6}
Y. Sekiwa, \emph{Thermodynamics of de Sitter black holes: thermal cosmological constant}, \emph{Phys. Rev.} {\bf D 73} (2006) 084009 [arXiv:hep-th/0602269].

\bibitem{7}
D. Kastor, S. Ray and J. Traschen, \emph{Enthalpy and the mechanics of AdS black holes}, \emph{Class. Quant. Grav.} {\bf 26} (2009) 195011 [arXiv:0904.2765].

\bibitem{8}
G. Gibbons, M. Perry, and C. Pope, \emph{The first law of thermodynamics for Kerr-anti-de Sitter black holes}, \emph{Class. Quant. Grav.} {\bf 22} (2005) 1503 [arXiv:hep-th/0408217].

\bibitem{9}
D. Kubiznak and R. B. Mann, \emph{P-V criticality of charged AdS black holes}, \emph{JHEP} {\bf 1207} (2012) 033 [arXiv:1205.0559].

\bibitem{10}
N. Altamirano, D. Kubiznak, R. B. Mann and Z. Sherkatghanad, \emph{Thermodynamics of rotating black holes and black rings: phase transitions and thermodynamic volume}, [arXiv:1401.2586].

\bibitem{11}
D. Kubiznak, R. B. Mann and M. Teo, \emph{Black hole chemistry: thermodynamics with Lambda}, [arXiv:1608.06147].

\bibitem{12}
John Lekner, \emph{Parametric solution of the van der Waals liquid-vapor coexistence curve}, \emph{Am. J. Phys.} {\bf 50} (1982) 161.

\bibitem{13}
K. Kornev, \emph{Lateral interactions of charges in thin liquid films and the Berezinskii-Kosterlitz-Thouless transition}, \emph{Phys. Rev.} {\bf E  60} (1999) 4371.

\bibitem{14}
G. Torrieri and I. Mishustin, \emph{Nuclear liquid-gas phase transition at large $N_{c}$ in the van der Waals approximation}, \emph{Phys. Rev.} {\bf C 82} (2010) 055202 [arXiv:10062471].

\bibitem{15}
M. N. Berberan-Santos, E. N. Bodunov and L. Pogliani, \emph{The van der Waals equation: analytical and approximate solutions}, \emph{J. Math. Chem.} {\bf 43} (2008) 1437 .

\bibitem{16}
D. C. Johnston, \emph{Thermodynamic properties of the van der Waals fluid}, [arXiv:1402.1205].

\bibitem{17}
S. W. Wei and Y. X. Liu, \emph{Clapeyron equations and fitting formula of the coexistence curve in the extended phase space of charged AdS black holes}, \emph{Phys. Rev.} {\bf D 91} (2015) 044018 [arXiv:1411.5749].

\bibitem{18}
S. W. Wei and Y. X. Liu, \emph{Insight into the microscopic structure of an AdS black hole from thermodynamical
phase transition}, \emph{Phys. Rev. Lett.} {\bf 115} (2015) 111302 [arXiv:1502.00386].







\end{thebibliography}
\end{document}